# Elucidation of charge storage characteristics of conducting polymer film using redox reaction

Asfiya Q. Contractor and Vinay A. Juvekar

Department of Chemical Engineering, Indian Institute of Technology Bombay, Powai, Mumbai 400 076, India

**Abstract:** A general technique to investigate charge storage characteristics of conducting polymer films has been developed. A redox reaction is conducted on a polymer film on a rotating disk electrode under potentiostatic condition so that the rate of charging of the film equals the rate of removal of the charge by the reaction. In an experiment on polyaniline film deposited on platinum substrate, using $Fe^{2+}/Fe^{3+}$ in HCl as the redox system, the voltammogram shows five distinct linear segments (bands) with discontinuity in the slope at specific transition potentials. These bands are the same as those indicated by ESR/Raman spectroscopy with comparable transition potentials. From the dependence of the slopes of the bands on concentration of ferrous and ferric ions, it was possible to estimate the energies of the charge carrier in different bands. It is shown that the charge storage in the film is capacitive.

**Key words:** conducting polymer, steady-state voltammetry, redox reaction, energy bands, capacitive charging, polyaniline.

**Introduction**

Understanding the charge storage characteristics of a conducting polymer is important from the point of view of its use as a capacitor. Commonly employed potentiodynamic techniques such voltammetry (step, pulse or cyclic) and impedance spectroscopy suffer from the disadvantage that the charging process involves several coupled dynamical steps. The overall dynamics is so complex that decoupling of the overall processes into individual steps is very difficult and can often be ambiguous. Many of these dynamical steps can be eliminated if charging can be performed under steady-state conditions. This can be achieved by conducting a suitable redox reaction in the film under potentiostatic condition, so that the rate at which the charge flows into the film equals the rate at which it is removed by the redox reaction. Since there is no charge accumulation in the polymer at steady state, there is no movement of counterions/ coions in and out of the polymer film to balance the excess charge. Also those charge relaxation processes (e.g. association of charges into polaron, bipolaron or their lattices) which occur at the rates faster than rate of the redox reaction, attain equilibrium at the steady state. Only three dynamical steps need to be accounted for, namely, transport of the redox species from the bulk solution to the film-solution interface, the surface reaction (which may be coupled with pore diffusion), and conduction of charge through the film. By conducting the reaction on a rotating disk electrode at different disk speeds, it is possible to eliminate the effect of mass transport using Levich-Koutecky plot. It is then possible to obtain only the contribution to current from the redox reaction. Moreover, the kinetics of the redox reaction gets coupled with the energy states of the charge carriers. Thus by monitoring the shift in the reaction kinetics as a function of the electrode potential, it is possible to gain an insight into the energy states of the charge carriers in different ranges of the electrode potential. Using this technique, it is possible to elucidate the energy band structure of the charge storage in polyaniline film, using $Fe^{2+}/Fe^{3+}$ and quinhydrone as the redox systems, as described here.

**Materials and methods**

All chemicals used in this study were A.R. grade. Aniline was redistilled under vacuum to obtain a colorless product. Ferrous ammonium sulphate and ferric ammonium sulphate were used as sources of ferrous and ferric ions respectively. Quinhydrone was 97% pure and was used without further purification.

All experiments were performed at the temperature of 298 K in a single compartment, three-electrode cell, with saturated calomel reference electrode and 5 mm diameter platinum rotating disk working electrode (Pine Instrument Company). The platinum electrode was polished using 0.05 μm alumina paste and was activated by potential cycling in 0.5M HCl. The film was deposited from a solution containing 0.1 M aniline and 0.5 M sulfuric acid, using potential cycling between -0.2V and 1.0V (disk speed 4000 RPM, sweep rate 50 mV/s), till the specified amount of the net charge (30±1 mC) was consumed in film synthesis. The redox charge of these films was measured to be 18±3 mC/cm$^2$ of the electrode area, which corresponded to a film thickness of about 1 $\mu m$, based on the criterion of Thyssen et al [1] (1 C/cm$^2$ redox charge $\equiv 47$ $\mu m$ film thickness).

Redox studies were performed in aqueous hydrochloric acid solution containing known concentrations of ferrous/ferric ions. The electrode potential was stepped down in small steps of 25 mV from 1 V to -0.2 V (film was irreversibly damaged when the potential was stepped up beyond 0.7 V; film characteristics were fully preserved in step-down mode). After each step-change, sufficient time was allowed for the system to attain steady state. The steady state current was recorded as a function of the electrode potential.

Redox currents in the absence of mass transfer were obtained by preparing Levich-Koutecky (LK) plots (plot of the reciprocal of the observed current density, $i$, versus inverse square root of the angular speed of the disc, $\omega$), which are based on the following equation and require that the redox reaction is first order in the concentrations of the redox species[2].

$$\frac{1}{i} = k\omega^{-1/2} + \frac{1}{i_R} \quad (1)$$

The first term on the right represents the contribution from mass transport of the redox species to the electrode surface. The term $i_R$ (reaction current density) is the contribution to the current density from the surface reaction (as well as other intra-film transport processes) and can be estimated from the intercept of the LK plot. Levich-Koutecky plots were found to be excellent straight lines and provided reliable estimates of the reaction current density. This also indicates that the redox reaction is first order with respect to the redox species.

**Results and Discussion:**

Figure 1(a) shows the plot of $i_R$ versus electrode potential, $V_e$ for the polyaniline film immersed in HCl solution containing $Fe^{2+}/Fe^{3+}$. Similar plot for quinhydrone is shown in Figure 1(b).

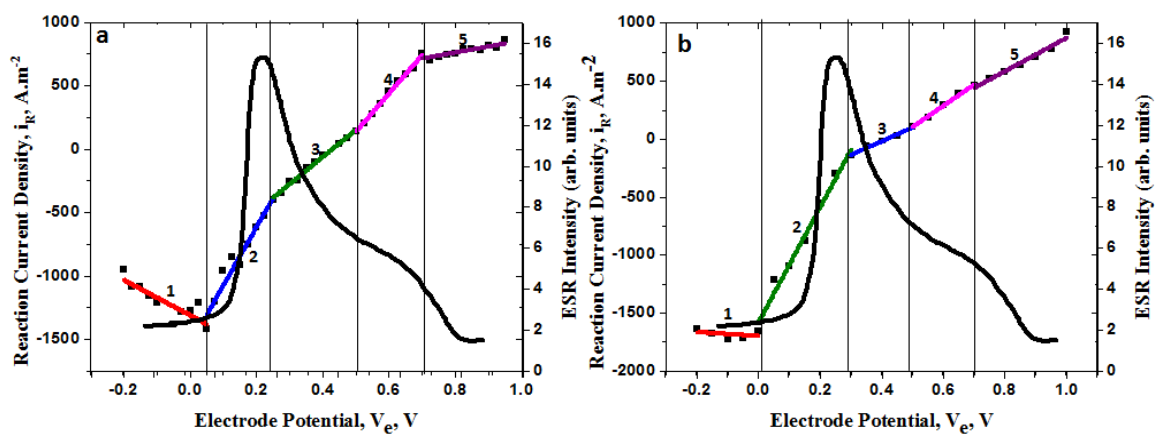

**Figure1.** Reduced voltammograms on 1 μm thick PANI film in 1.0 M. HCl having 18.9mM $Fe^{2+}$ + 9.4mM $Fe^{3+}$ (1-a)and 10mM Quinhydrone in 0.5M HCl(1-b). Five different bands are seen in the form of distinct straight lines, with transition potentials indicated by vertical lines. Both plots are overlaid with an esrogram of PANI film in 1.0 M HCl (adapted from reference 3). The bands are numbered 1 to 5. (All potentials are relative to saturated calomel electrode)

The voltammograms show five linear segments, each having a distinct slope. Transition from one segment to the next is sharp indicating discontinuity in the first derivative $di_R/dV_e$. These straight lines are numbered on the plot from 1 to 5. Transition from one line to the next occurs at definite electrode potentials (within ±0.02 V). These potentials are independent of the concentrations of the redox species and their values, in increasing order, are: 0.05, 0.30, 0.50 and 0.70 V.

Electron spin resonance spectroscopy [3] as well as Raman spectroscopy [4] data indicate existence of five energy bands in polyaniline, corresponding to different structures of the polyaniline chains. To see the correspondence between our data and the spectroscopic data, an esrogram of PANI film is superimposed on Figure 1. Transitions seen in our data are comparable to those seen in the esrogram. We therefore name the bands 1 to 5 by leucoemaraldine (zero spin), polarons (increase in electron spin with potential), polaron lattice (rapid decrease in spin), bipolaron (slow decrease in spin) and pernigraniline (zero spin).

It is seen from Figure 1 that the linear trend persists even when crossover from the cathodic to the anodic region occurs. This indicates the absence of charge transfer resistance. We therefore propose that the charging process is capacitive. We explain this process as follows.

When the electrode potential is increased, charges are injected into the film, converting the undoped polymer sites $P$ into doped sites $P^+$. These sites react with the redox species according to the reaction $Fe^{2+} + P^+ = Fe^{3+} + P$. At steady state, the current flowing from the electrode into the film equals that consumed by the redox reaction. Hence assuming the reaction to be reversible and second order, we can express $di_R/dV_e$ (in each band) in terms of the concentrations of redox species in the solution ($[Fe^{2+}]$ and $[Fe^{3+}]$) and the surface densities of undoped and doped sites on the polymer surface ($[P]$ and $[P^+]$ respectively):

$$\frac{di_R}{dV_e} = Fa_f \left( k_f \frac{d[P^+]}{dV_e}[Fe^{2+}] - k_b \frac{d[P]}{dV_e}[Fe^{3+}] \right) \quad (2)$$

Here, $k_f$ and $k_b$ are the rate constants for the forward and the backward surface reactions, respectively, and $a_f$ is the surface area of the polymer per unit area of the electrode. Since the reaction is non-electrochemical, the rate constants are not functions of the electrode potential and hence can be treated as constants in a given band.

If the surface density of the total sites (doped plus undoped) is assumed to remain constant at $[S]$, so that $[P]+[P^+]=[S]$, we can replace $d[P]/dV_e$ by $-d[P^+]/dV_e$. We can also see from Eq 2 that since in a given band $di_R/dV_e$ is constant, so is $d[P^+]/dV_e$, indicating that charging is a linear capacitive process. We denote the differential capacitance in band $j$, by $c_{fj}$ and write

$$F\left(\frac{d[P^+]}{dV_e}\right)_j = c_{fj}, \; j=1,2...,5 \quad (3)$$

Eq 2 can therefore be written as

$$\left(\frac{di_R}{dV_e}\right)_j = a_f c_{fj}\left(k_{f,j}[Fe^{2+}] + k_{b,j}[Fe^{3+}]\right) \quad (4)$$

Eq 4 implies that $di_R/dV_e$ should vary linearly with concentrations of ferrous and ferric ions, with slopes $a_f c_{f,j} k_{fj}$ and $a_f c_{fj} k_{bj}$, respectively. The linear dependence is seen in Figure-2 (a) (for $Fe^{2+}$) and Figure-2 (b) (for $Fe^{3+}$)

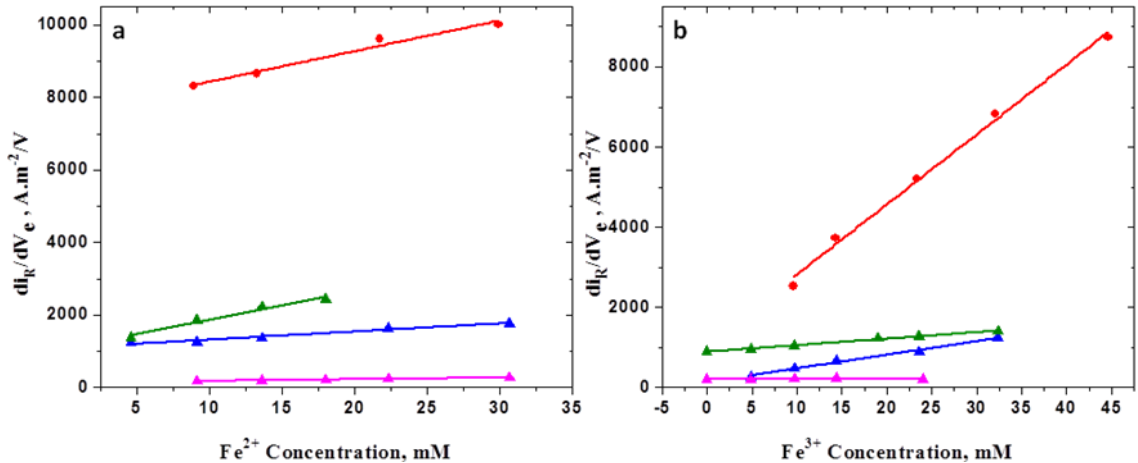

**Figure2. Effect of ferrous ions (1-a) and ferric ions (1-b) on the slope of the line segments ( $di_R/dV_e$ ) of the voltammograms. Voltammograms are prepared on 1μm PANI film in 0.5M HCl. Key: Red (Polaron band); Blue (Polaron lattice band); Green (Bipolaron band); Pink (Pernigraniline band). (All potentials are relative to saturated calomel electrode)**

The values of the slopes are listed in Table-1. A large variation in the slopes is observed from band-to-band. The ratio of the slopes ($a_f c_{f,j} k_{fj}$ and $a_f c_{fj} k_{bj}$) in $j^{th}$ band gives the equilibrium constant $K_{Tj} = k_{fj}/k_{bj}$ for the reaction $Fe^{2+} + P^+ = Fe^{3+} + P$, from which the standard free energy for reaction in band $j$ ($\Delta G_{Tj}^0 = -RT \ln K_j$) can be computed. Both $K_{Tj}$ and $\Delta G_{Tj}^0$ are listed in Table-1.

**Table 1:** Characteristics of the energy bands of Polyaniline (PANI) film.

| Polyaniline Structure[3] | Potential Range V | $a_f c_{fj} k_{fj}$ | $a_f c_{fj} k_{bj}$ | $K_{Tj} = k_{f_j}/k_{b_j}$ | $\Delta G_{Tj}^0$ kJ.mol$^{-1}$ | $\Delta G_{Cj}^0$ kJ.mol$^{-1}$ | $K_{Cj} \times 10^9$ |
|---|---|---|---|---|---|---|---|
| Polaron | 0.05-0.3 | 84.3 | 175 | 0.483 | 1.80 | 41.91 | 45.0 |
| Polaron lattice | 0.3-0.5 | 22.4 | 33.8 | 0.663 | 1.02 | 42.89 | 30.3 |
| Bipolaron | 0.5-0.7 | 78.9 | 16.2 | 4.86 | -3.92 | 47.63 | 4.48 |
| Pernigraniline | 0.7-1.0 | 4.7 | 0.217 | 21.9 | -7.65 | 51.36 | 0.993 |

We can split the overall reaction into two coupled steps: $Fe^{2+} = Fe^{3+} + e^- (\Delta G_R^0)$ and $P = P^+ + e^- (\Delta G_{Cj}^0)$ so that $\Delta G_{Tj}^0 = \Delta G_R^0 - \Delta G_{Cj}^0$. The standard free energy $\Delta G_R^0$ is independent of the nature of the electrode. It is estimated from the formal potential (0.453 V in 0.5 M HCl both on PANI and platinum electrode) as 43.71 kJ.mol$^{-1}$. Knowing $\Delta G_{Tj}^0$, the value of $\Delta G_{Cj}^0$, the free energy for charging, along with the corresponding equilibrium constant $K_{Cj}$, can be estimated. These values are listed in Table-1. It is seen that $K_{Cj}$ decreases by a factor of 50 as we move from polaron band to pernigraniline band. This means the energy of the charge carrier increases from the polaron to pernigraniline band. Difference in energies between polaron lattice and polaron is small but the difference between bipolaron and polaron lattice is large and nearly equal to that between pernigraniline and bipolaron.

Steady-state voltammograms in Leucoemaraldine band are shown in Figure 3(a). Unlike the other bands, the slope $di_R/dV_e$ in this band is insensitive to concentration of ferrous and ferric ions. Also since no anodic current is observed over a wide range of ferrous ion concentrations, we conclude that leucoemaraldine band is devoid of $P^+$. However, increase in the concentration of ferric ions produces a parallel downward shift of the polarization curves. This point is emphasized further in Figure 3(b), where the current density at a fixed potential is plotted against concentration of ferrous and ferric ions. For ferric ions, the plot is a straight line passing through origin. From this, we propose that the current density in leucoemaraldine band is given by $i_R = -Fa_f k_{b_1}[S][Fe^{3+}]$. Thus ferric ions react with the uncharged sites in the film and generate $P^+$ which are completely neutralized by the flow of electrons from the substrate electrode, resulting in cathodic current.

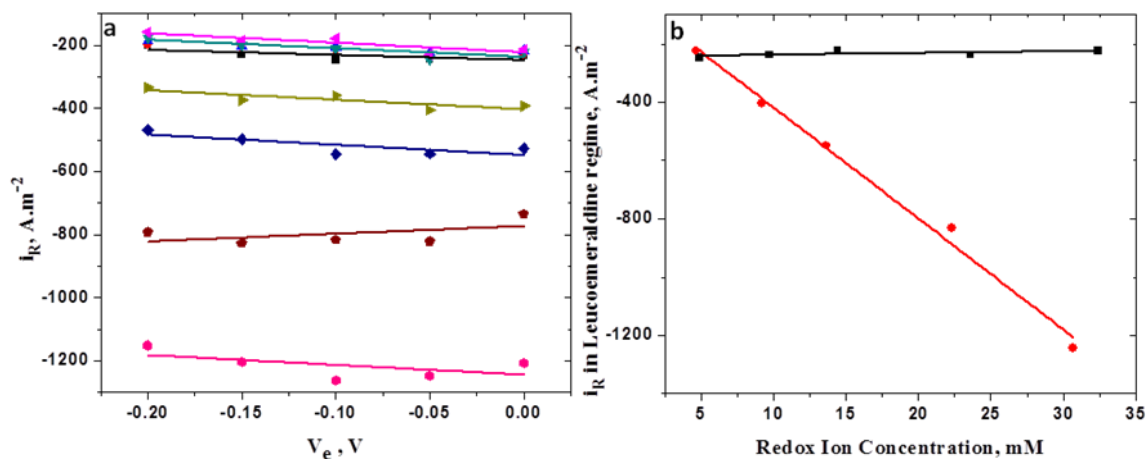

**Figure 3. Steady-state voltammograms on 1μm thick PANI Film in leucoemeraldine band (3a) and dependance of the reaction current density on redox ion concentrations (3b)**

**Legend Colour:3(a)-Black ([Fe$^{2+}$] =4.9mM, [Fe$^{3+}$] =4.9mM); Dark Cyan (23.58mM, 4.72mM); Magenta (32.41mM, 4.63mM); Dark Yellow (32.11mM, 9.17mM); Navy Blue (31.82mM, 13.64mM); Brown (31.25mM, 22.32mM); Pink (30.70mM, 30.70mM) ; 3(b)- Black ([Fe$^{2+}$]) and Red ([Fe$^{3+}$]).**

**Conclusions**

The present method allows us to study charge storage characteristics of conducting polymer films under steady dynamic conditions. Great simplification is achieved since the dynamics of double layer and the associated counterion diffusion can be eliminated. Since the rate of charge transport and relaxation must match with the rate of reaction, it is possible to investigate charge relaxation steps of different time constants by choosing redox reactions with appropriate rate constants. Specifically for PANI films, the present study has revealed the band structure of charge storage and the associated band energies. It was also observed that the charge transfer at the polymer film/substrate electrode interface is not rate controlling. Our method is general and can be applied to any intrinsically conducting polymer.